\documentclass[prl,preprint,aps,amsmath,amssymb,longbibliography,floatfix]{revtex4-2}

\usepackage{graphicx}
\usepackage{amsmath,amssymb,bm,epsfig}
\usepackage{dcolumn}
\usepackage{bm}

\graphicspath{ {./Figures/} }

\begin{document}

\preprint{Phys. Rev. B 112, 245430 (2025)}

\title{Purcell-enhanced quantum adsorption}

\author{Dennis P. Clougherty}

\affiliation{
Department of Physics\\
University of Vermont\\
Burlington, VT 05405-0125}

\date{April 14, 2025}

\begin{abstract}
Cold atoms can adsorb to a surface with the emission of a single phonon when the binding energy is sufficiently small.   The effects of phonon damping and adsorbent size on the adsorption rate in this quantum regime are studied using the multimode Rabi model.  It is demonstrated that the adsorption rate can be either enhanced or suppressed relative to the Fermi golden rule rate, in analogy to cavity effects in the spontaneous emission rate in QED.  A mesoscopic-sized adsorbent behaves as an acoustic cavity that enhances the adsorption rate when tuned to the adsorption transition frequency and suppresses the rate when detuned.  This acoustic cavity effect occurs in the regime where the frequency spacing between vibrational modes exceeds the phonon linewidth.  
 \end{abstract}

\maketitle
 
\newpage

\section{Introduction}

The radiative properties of an atom can be radically altered by placing the atom in a cavity.  If the cavity is tuned to an atomic transition frequency, the spontaneous emission rate can be enhanced relative to its free-space value; if the cavity is detuned from the transition frequency, spontaneous emission can be suppressed.  This Purcell effect \cite{purcell} is a manifestation of the interaction of the atom with the cavity-modified electromagnetic vacuum. 

In solids, there is an analogous acoustic Purcell effect; e.g.,  a color-center in diamond can function as an excited atom, decaying with the emission of phonons (real and virtual).  By placing the color center in a nanomechanical resonator tuned to the spin transition frequency, the spin relaxation rate can be enhanced \cite{acoustic-purcell} by a factor of 10.

There are a number of effects in solids that involve the emission or absorption of phonons (e.g., thermal and electric conduction, BCS superconductivity, and optical absorption and emission).  The acoustic Purcell effect may well be used to shape the properties of solids by controlling interactions with phonons.  Mechanical metamaterials \cite{acoustic-meta} where the phonon properties of the material can be manipulated may provide a way to achieve this experimentally.

In this work, the effect of an acoustic cavity on the adsorption of a cold adsorbate is studied.  Using the Dirac-Frenkel variational principle \cite{dirac, frenkel}, a time-dependent description of phonon-assisted quantum adsorption is formulated, and a closed-form expression for the acoustic Purcell effect on the phonon-assisted adsorption rate is obtained.  The finite size of the adsorbent creates an acoustic cavity that modifies the density of vibrational modes, alters the adsorbate-phonon interaction, and consequently enhances or suppresses the adsorption rate.  

\section{Multimode quantum Rabi model}

The multimode quantum Rabi Hamiltonian \cite{mqrm,rabi-model} gives a simplified model for describing phonon-assisted quantum adsorption on a 2D adsorbent \cite{dpc24}
\begin{equation}
H=H_a+H_p+H_{i}
\end{equation}
where
\begin{eqnarray}
H_a&=& E_c c^\dagger c -E_b b^\dagger b\\
H_p&=& \sum_n \omega_n a^\dagger_n a_n\\
H_{i}&=&-g(c^\dagger b+b^\dagger c) \sum_n  (a^\dagger_n +a_n)
\label{model}
\end{eqnarray} 

The model considers two states of the adsorbate: the first is the initial state of the adsorbate in the gas phase with energy $E_c$; the second is the adsorbate bound to the surface with energy $-E_b$.  $c^\dagger$ ($c$) creates (annihilates) an adsorbate in the gas phase, while $b^\dagger$ ($b$) creates (annihilates) an adsorbate bound to the surface. $a_n^\dagger$ ($a_n$) creates (annihilates) a phonon in the nth mode.   Adsorption occurs by displacement of the adsorbent which is assumed to be an elastic membrane under tension.  

The coupling constant $g$ is a matrix element of the normal derivative of the static surface potential, as described in Ref.~\cite{dpc13}. For cold adsorbates, $g$ varies as the square root of the adsorbate energy in the gas phase \cite{sengupta} and can be experimentally controlled \cite{yu93}. 

The adsorbent is taken to be a disk of radius $a$ that is clamped at its edge, creating an acoustic resonator.  Acoustic resonators of this type have been fabricated by suspending graphene over pores on the surface of a SiO$_2$ substrate \cite{mceuen} using mechanical exfoliation.  The vibrational spectrum of circularly symmetric modes is taken to be $\omega_n= c\pi n/a$ ($n=1\dots N$) where $c$ is the transverse speed of sound.  (This approximation for the frequency spacing becomes asymptotically exact for $n\gg 1$.)

\section{Variational ansatz}
Following Ref.~\cite{dpc24}, a time-dependent description of the adsorption dynamics can be formulated with the application of the Dirac-Frenkel variational principle \cite{dirac, frenkel}.
A time-dependent variational state that describes the fundamental adsorption process is chosen, and an effective Lagrangian is obtained for the system.   Time-dependent amplitudes in the variational state serve as generalized coordinates.  Equations of motion for the variational amplitudes follow from the Euler-Lagrange equations for the effective Lagrangian.   The equations of motion can be subsequently solved using integral transform methods.

The variational state of the system is taken to be a superposition of two types of states: the initial state of the adsorbate of energy $E_c$ with a thermal distribution of phonons supported on the adsorbent; and secondly, the adsorbate bound to the adsorbent with an additional phonon present.  

\begin{equation}
|\psi(t)\rangle=\bigg(C(t) c^\dagger+\sum_m B_m(t) a_m^\dagger b^\dagger\bigg) |{\rm \{n_q\}}\rangle
\label{psi}
\end{equation}
where $|{\rm \{n_q\}}\rangle=\prod_q \frac{(a_q^\dagger)^{n_q}}{\sqrt{n_q!}}|0\rangle$. $C(t)$ and the set of $B_n(t)$ are taken to be variational functions.  

The effective Lagrangian in the Dirac-Frenkel approach is given by
\begin{equation}
L=\langle\psi(t)| \bigg(i\frac{d}{dt} - H\bigg)|\psi(t)\rangle
\end{equation}
For the variational ansatz in Eq.~\ref{psi}, the following Lagrangian results after thermal-averaging the phonon matrix elements
\begin{eqnarray}
L&=&iC^*\frac{dC}{dt}+i\sum_p (n_p+1)B_p^*\frac{dB_p}{dt}-(E_c+\sum_m n_m\omega_m) C^* C\nonumber\\
&+&
\sum_m B_m^* B_m(n_m+1)(E_b-\sum_p n_p\omega_p)+g\sum_m(C^*B_m+B_m^*C)(n_m+1)
\label{lagrangian}
\end{eqnarray}
where $n_m={1/(\exp(\omega_m/T)-1)}$, the initial thermal distribution (mode $m$) of phonons in the adsorbent.

The Euler-Lagrange equations for this Lagrangian are

\begin{eqnarray}
\label{euler-lagrange1}
i\frac{dC}{dt}&=&(E_c+\sum_p n_p\omega_p)C-{g}\sum_m (n_m+1)B_m\\
i\frac{dB_n}{dt}&=&-(E_b-\omega_n-\sum_p n_p\omega_p)B_n-{g} C
\label{euler-lagrange2}
\end{eqnarray}
  
The dynamics is found in the solution of this set of coupled linear first-order equations for the variational functions subject to the initial conditions that the adsorbate starts in the gas phase at $t=0$ ($C(0)=1$ and $B_n(0)=0$).  

Eqs.~\ref{euler-lagrange1} and \ref{euler-lagrange2} can be solved analytically with Laplace transforms; for example, the Laplace-transformed amplitude for the entrance channel ${\tilde C(s)}$ is   
\begin{equation}
{\tilde C(s)}={\frac{i}{i s-E_c -\sum_p n_p\omega_p-g^2\sum_p\frac{n_p+1}{i s+E_b-\omega_p-\sum_m n_m\omega_m}}}
\label{C(s)}
\end{equation}
The time-dependent amplitudes $C(t)$, $B_m(t)$ can be obtained by inverse transforming using the Bromwich contour in the complex s-plane.  
 
The poles of ${\tilde C(s)}$ are $N+1$ solutions to 
\begin{equation}
i s_n-E_c -\sum_p n_p\omega_p-g^2 \Sigma(i s_n)=0
\end{equation}
where 
\begin{equation}
\Sigma(i s)=\sum_p\frac{n_p+1}{i s+E_b-\omega_p-\sum_m n_m\omega_m}
\end{equation}

\section{Adsorption rate}

The adsorption rate may be obtained from the adsorbate self-energy using 
\begin{equation}
R\approx -2 g^2\ {\rm Im}\ \Sigma(E)
\label{sumR}
\end{equation}
where $E=E_c+\sum_p n_p\omega_p$, the initial energy of the system. 
In the absence of phonon damping, there is no true adsorption for a finite-size adsorbent and $R=0$. There are only two possibilities for the adsorbate: prompt elastic scattering back to the gas phase or the excitation of a resonance \cite{dpc92} that decays back to the gas phase after a time delay.   
 
 For a suitably large number of modes, the sum implicit in Eq.~\ref{sumR} may be replaced by an integral over phonon frequency $\omega$ in the quasicontinuum approximation
 \begin{equation}
 R_0\approx -2 g^2\ {\rm Im}\ \int_0^{\omega_D} d\omega {\cal D}_0\frac{n(\omega)+1}{E_c+E_b-\omega}
 \end{equation}
 ${\cal D}_0$ is the vibrational density of circularly symmetric modes.
 
 This integral as it stands is ambiguous, as a singularity lies on the integration path; however, the inevitable clamping loss that damps the phonons suggests that physically the phonon frequencies must acquire a small imaginary part $\omega\to \omega-i \eta$.  This resolves the integral ambiguity and gives the following adsorption rate in the limit $\eta\to 0^+$  
\begin{equation}
R_0= 2\pi g^2{\cal D}_0 {(n(\Omega_s)+1)}\Theta(\omega_D-\Omega_s)
\label{R}
\end{equation}
where $\Omega_s=E_c+E_b$, $\omega_D$ is the highest vibrational frequency supported by the membrane, and $\Theta$ is the Heaviside function.  This is the same result obtained with Fermi's golden rule.
 
 \begin{figure}[htbp]
\includegraphics[width=8cm]{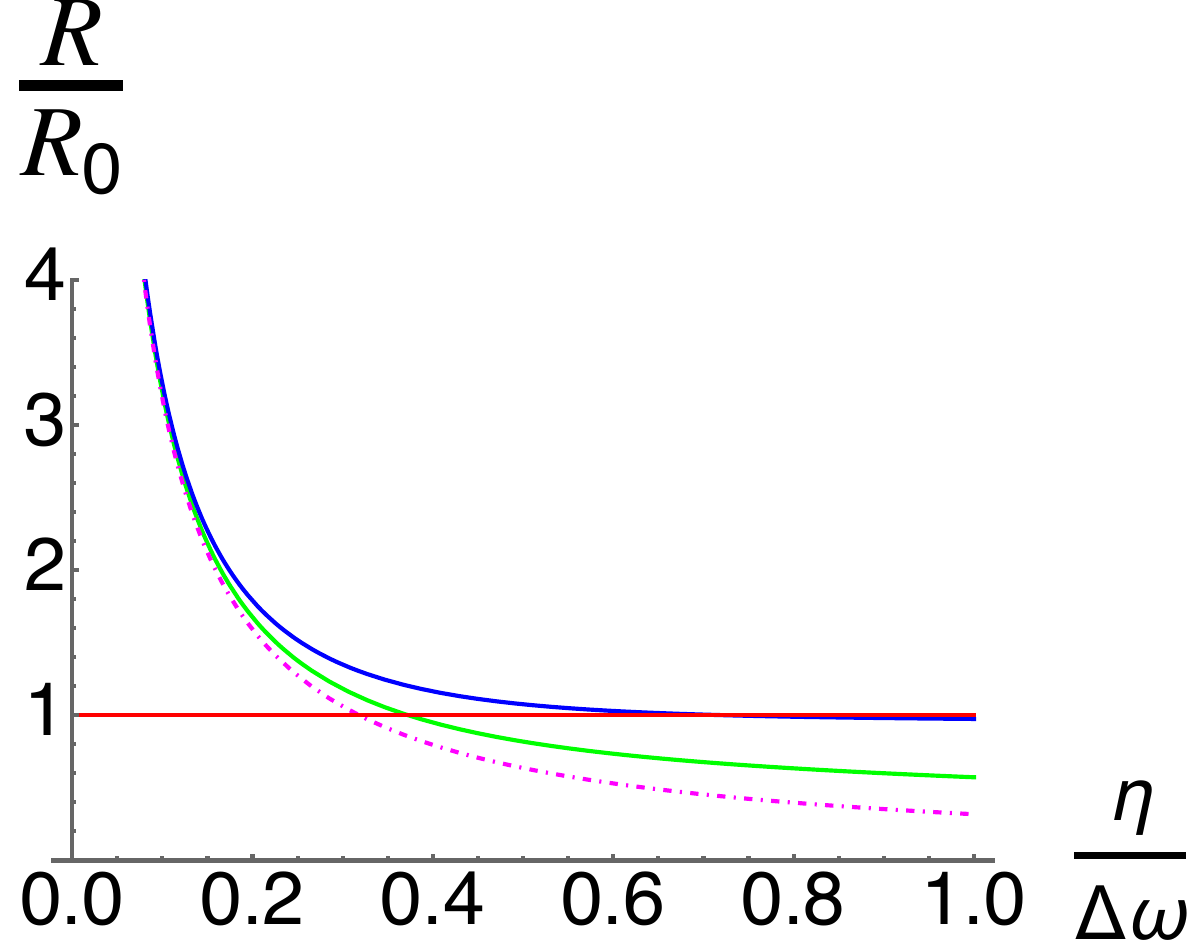}
\caption{\label{fig:on-res} Relative adsorption rate $R/R_0$ versus coupling strength $\frac{\eta}{\Delta\omega}$ for on-resonance cases $N=4$ and $N_s=1$ (green);  $N=60$ and $N_s=15$ (blue).  The leading asymptotic behavior $R/R_0$ for $\frac{\eta}{\Delta\omega}\to 0$ given in Eq.~\ref{asymp} is plotted (magneta, dot-dashed).   Fermi golden rule result (horizontal, red) is displayed for comparison.  The rate is enhanced relative to $R_0$  for low $\frac{\eta}{\Delta\omega}$. }
\end{figure}

However, for a mesoscopic adsorbent, the quasicontinuum approximation to the sum can be a poor approximation.   For low temperature $T\ll \Delta\omega$, the adsorption rate can be rewritten
 \begin{eqnarray}
 R&\approx& -2 g^2 \ {\rm Im}\ \sum_{m=1}^N \frac{1}{(\Omega_s-\omega_m+i\eta)}\\
 &=&  \frac{2 g^2}{\Delta\omega}\ {\rm Im}\ \bigg( \psi\big(N_s+i \frac{\eta}{\Delta\omega}-N\big)-\psi\big(N_s+i \frac{\eta}{\Delta\omega}\big)\bigg)\nonumber
 \end{eqnarray}
 where $\psi(z)$ is the digamma function \cite{g&r} and  $N_s\equiv\frac{a \Omega_s}{\pi c}$.  ($N_s$ is restricted to be less than $N$ so that adsorption by single phonon emission is energetically possible.)
 
 The adsorption rate relative to $R_0$ is independent of $g$ and is given by
  \begin{equation}
 \frac{R}{R_0}
 =  \frac{1}{\pi}\ {\rm Im}\ \bigg( \psi\big(N_s+i \frac{\eta}{\Delta\omega}-N\big)-\psi\big(N_s+i \frac{\eta}{\Delta\omega}\big)\bigg)
\label{RR0}
 \end{equation}
 (For $\eta\ll\Omega_s$, this expression is also valid in the high temperature regime.)

$\psi(z)$ has simple poles at the negative integers \cite{g&r}.  Consequently, 
the asymptotic behavior of $R/R_0$ as $\frac{\eta}{\Delta\omega}\to 0$ depends sensitively on $N_s$, viz.
 \begin{equation}
 \frac{R}{R_0}\sim \begin{cases}
			 \frac{\Delta\omega}{\pi\eta}, & \text{$N_s\in\mathbb{N}^+$}\\
            \frac{1}{\pi}(\zeta(2,N_s-N)-\zeta(2,N_s))\frac{\eta}{\Delta\omega}, & \text{otherwise}
		 \end{cases}
\label{asymp}
\end{equation}
where $\zeta(s,q)$ is the Hurwitz zeta function.

\begin{figure}[htbp]
\includegraphics[width=8cm]{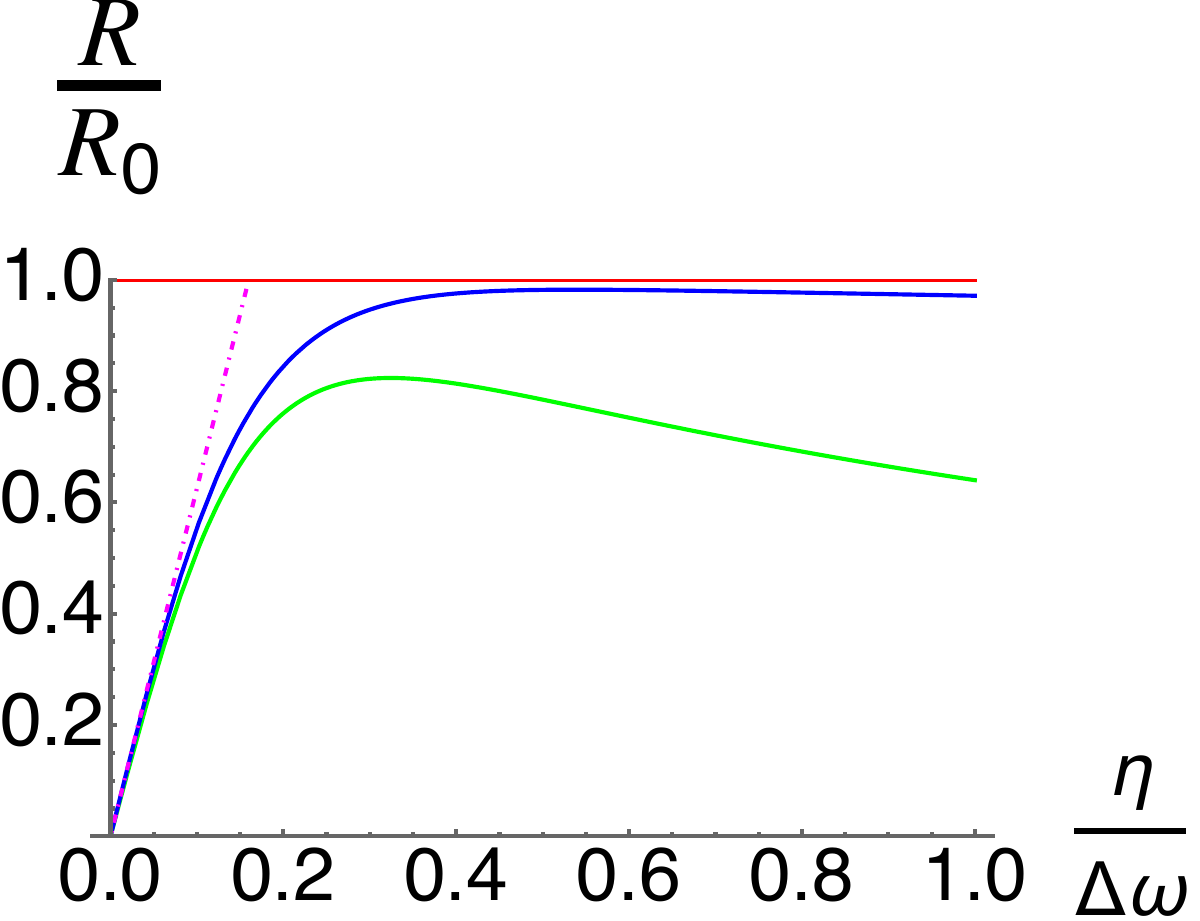}
\caption{\label{fig:off-res} Relative adsorption rate $R/R_0$ versus coupling strength $\frac{\eta}{\Delta\omega}$ for  off-resonance cases $N=60$ with $N_s=14.25$ (blue); $N=5$ with $N_s=1.25$ (green).  The leading asymptotic behavior $R/R_0$ for $\frac{\eta}{\Delta\omega}\to 0$ is plotted (magneta, dot-dashed).   Purcell suppression is manifest for all $\frac{\eta}{\Delta\omega}$.}
\end{figure}

 If $N_s\in\mathbb{N}^+$, then there is an enhancement in the relative adsorption rate (see Fig.~\ref{fig:on-res}).  An adsorption rate enhancement by a factor of ten would then require a phonon damping rate of $\eta\sim \Delta\omega/10\pi$.

\begin{figure}[htbp]
\includegraphics[width=8cm]{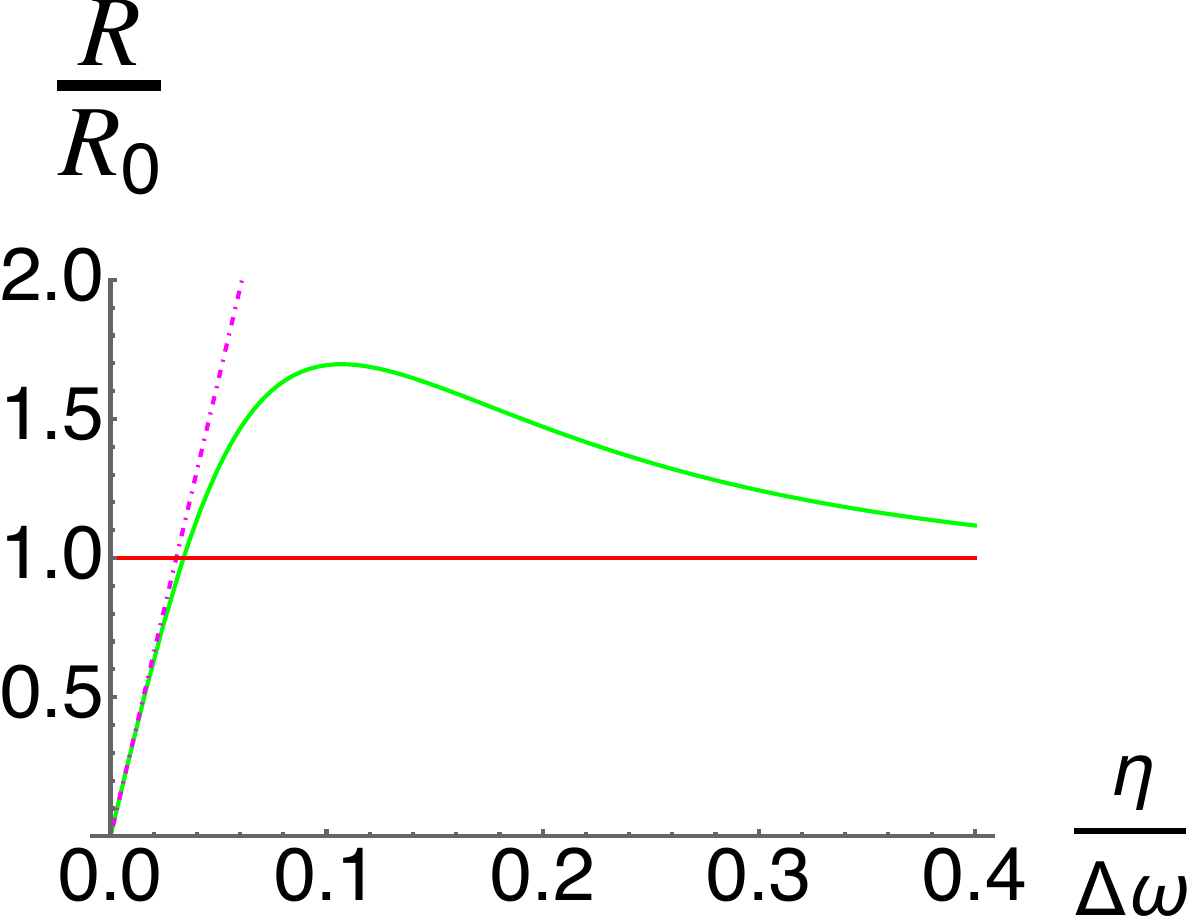}
\caption{\label{fig:near-res} Relative adsorption rate $R/R_0$ versus coupling strength $\frac{\eta}{\Delta\omega}$ for a near-resonance case $N=40$  with $N_s=9.89$ (green).  The leading asymptotic behavior $R/R_0$ for $\frac{\eta}{\Delta\omega}\to 0$ is plotted (magneta, dot-dashed).   Purcell enhancement is evident for $\frac{\eta}{\Delta\omega}\gtrsim 0.04$, while suppression occurs below this threshold.}
\end{figure}

However, if $N_s\notin\mathbb{N}^+$, then there is complete suppression of the adsorption rate for $\eta=0^+$ (see Figs.~\ref{fig:off-res} and \ref{fig:near-res}).

Fortunately, even if $N_s$ is close to an integer,  there can be an intermediate regime where there is an enhancement of the relative adsorption rate above a finite damping threshold $\eta_0$.  Figure \ref{fig:near-res} provides an example of this for $N_s=9.89$ with $\eta_0\approx 0.04\ \Delta\omega$.

A slight modification of the experiment by Yu et al. \cite{yu93} may lead to an experimental test of  Purcell-enhancement of quantum adsorption.  Yu et al. \cite{yu93} measured the adsorption of cold hydrogen atoms on a liquid helium film.  A gas of hydrogen atoms was magnetically trapped and cooled to submillikelvin temperatures in a cell.  A copper disk supported the liquid helium film at the bottom of the cell.  The confining field is shut off and the gas comes in contact with the surface of the helium film.  Hydrogen atoms may be adsorbed with the creation of a ripplon on the surface of the liquid helium film.  Yu et al. \cite{yu93} measured the adsorption rate on films of varying thicknesses for a range of gas temperatures.  

\begin{figure}[htbp]
\includegraphics[width=8cm]{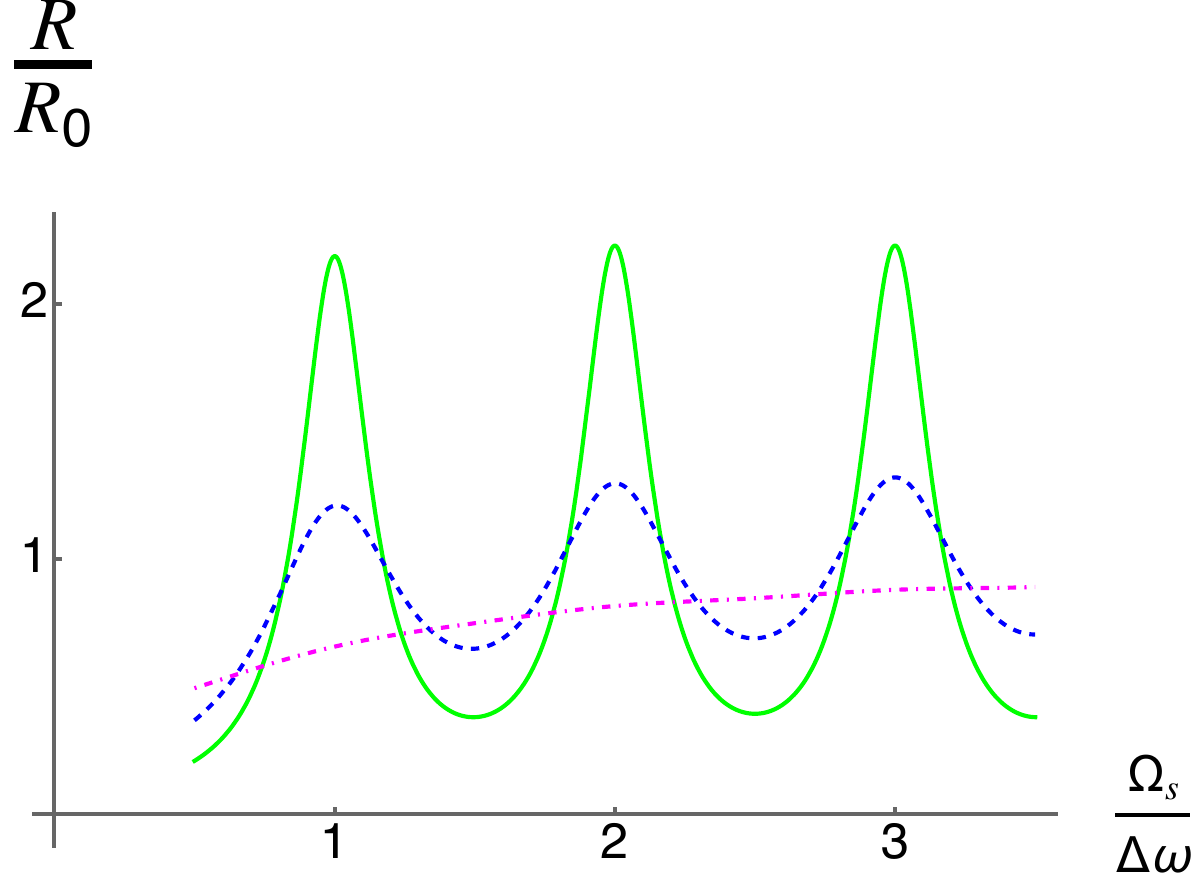}
\caption{\label{fig:sweep} Relative adsorption rate $R/R_0$ displays oscillations with adsorbate energy $E_c$ for a finite-size adsorbent [$\frac{\eta}{\Delta\omega}=0.15$  (green); $\frac{\eta}{\Delta\omega}=0.3$ (blue, dashed), $\frac{\eta}{\Delta\omega}=1$ (magenta, dot-dashed)].  For nonlinear dispersion, the frequency spacing between successive modes is not constant, but it may be nearly so over a sufficiently small energy interval.}
\end{figure}

By etching and cesiating the substrate surface to produce a dense array of pores \cite{nanoporous}, puddles of liquid helium could be formed in the presence of the surface disorder \cite{puddles3} to create ripplon cavities that would have an enhanced adsorption rate for resonant frequency ripplons.  The measured adsorption rate would then display oscillations with varying gas temperature (see Fig.~\ref{fig:sweep}).  With the addition of helium to the cell, puddles would grow and merge to form a thin film of helium, and the oscillations would disappear.  The variation of the relative adsorption rate $R/R_0$ with $\Omega_s$ for a porous surface is given in Fig.~\ref{fig:sweep}.  The frequency spacing $\Delta\omega$ varies  inversely with the puddle size for excitations with a linear dispersion.  For ripplons, $\omega\propto q^{\frac{3}{2}}$.  Thus, the frequency spacing would only be approximately constant over a sufficiently small energy interval.

These results  provide an example of a cavity-enhanced reaction, an acoustic analog of the work in polariton chemistry \cite{polariton}.  One direct application of this effect is found in the fabrication of quantum devices: the creation of acoustic cavities on the surface of an adsorbent using architected mechanical metamaterials \cite{mech-mat} might be used to modify the adsorption rate of cold adsorbates.  For quantum devices such as atom chips \cite{chip04} or atom mirrors, this could be a way to suppress unwanted adsorption \cite{sadeghpour} that degrades device performance.   

Analogous to the experimental studies in cavity QED \cite{haroche}, the acoustic Purcell effect may offer a way to probe the phonon vacuum and gain insight into properties of condensed matter systems at the mesoscopic scale.  Support of this work under NASA grant number 80NSSC19M0143 is gratefully acknowledged.  

\bibliography{purcell}

\begin{thebibliography}{21}%
\makeatletter
\providecommand \@ifxundefined [1]{%
 \@ifx{#1\undefined}
}%
\providecommand \@ifnum [1]{%
 \ifnum #1\expandafter \@firstoftwo
 \else \expandafter \@secondoftwo
 \fi
}%
\providecommand \@ifx [1]{%
 \ifx #1\expandafter \@firstoftwo
 \else \expandafter \@secondoftwo
 \fi
}%
\providecommand \natexlab [1]{#1}%
\providecommand \enquote  [1]{``#1''}%
\providecommand \bibnamefont  [1]{#1}%
\providecommand \bibfnamefont [1]{#1}%
\providecommand \citenamefont [1]{#1}%
\providecommand \href@noop [0]{\@secondoftwo}%
\providecommand \href [0]{\begingroup \@sanitize@url \@href}%
\providecommand \@href[1]{\@@startlink{#1}\@@href}%
\providecommand \@@href[1]{\endgroup#1\@@endlink}%
\providecommand \@sanitize@url [0]{\catcode `\\12\catcode `\$12\catcode
  `\&12\catcode `\#12\catcode `\^12\catcode `\_12\catcode `\%12\relax}%
\providecommand \@@startlink[1]{}%
\providecommand \@@endlink[0]{}%
\providecommand \url  [0]{\begingroup\@sanitize@url \@url }%
\providecommand \@url [1]{\endgroup\@href {#1}{\urlprefix }}%
\providecommand \urlprefix  [0]{URL }%
\providecommand \Eprint [0]{\href }%
\providecommand \doibase [0]{https://doi.org/}%
\providecommand \selectlanguage [0]{\@gobble}%
\providecommand \bibinfo  [0]{\@secondoftwo}%
\providecommand \bibfield  [0]{\@secondoftwo}%
\providecommand \translation [1]{[#1]}%
\providecommand \BibitemOpen [0]{}%
\providecommand \bibitemStop [0]{}%
\providecommand \bibitemNoStop [0]{.\EOS\space}%
\providecommand \EOS [0]{\spacefactor3000\relax}%
\providecommand \BibitemShut  [1]{\csname bibitem#1\endcsname}%
\let\auto@bib@innerbib\@empty
\bibitem [{\citenamefont {Purcell}(1946)}]{purcell}%
  \BibitemOpen
  \bibfield  {author} {\bibinfo {author} {\bibfnamefont {E.~M.}\ \bibnamefont
  {Purcell}},\ }\bibfield  {title} {\bibinfo {title} {Spontaneous emission
  probabilities at radio frequencies},\ }\href@noop {} {\bibfield  {journal}
  {\bibinfo  {journal} {Phys. Rev.}\ }\textbf {\bibinfo {volume} {69}},\
  \bibinfo {pages} {681} (\bibinfo {year} {1946})}\BibitemShut {NoStop}%
\bibitem [{\citenamefont {Joe}\ \emph {et~al.}(2025)\citenamefont {Joe},
  \citenamefont {Haas}, \citenamefont {Kuruma}, \citenamefont {Jin},
  \citenamefont {Kang}, \citenamefont {Ding}, \citenamefont {Chia},
  \citenamefont {Warner}, \citenamefont {Pingault}, \citenamefont {Machielse},
  \citenamefont {Meesala},\ and\ \citenamefont {Loncar}}]{acoustic-purcell}%
  \BibitemOpen
  \bibfield  {author} {\bibinfo {author} {\bibfnamefont {G.}~\bibnamefont
  {Joe}}, \bibinfo {author} {\bibfnamefont {M.}~\bibnamefont {Haas}}, \bibinfo
  {author} {\bibfnamefont {K.}~\bibnamefont {Kuruma}}, \bibinfo {author}
  {\bibfnamefont {C.}~\bibnamefont {Jin}}, \bibinfo {author} {\bibfnamefont
  {D.~D.}\ \bibnamefont {Kang}}, \bibinfo {author} {\bibfnamefont
  {S.}~\bibnamefont {Ding}}, \bibinfo {author} {\bibfnamefont {C.}~\bibnamefont
  {Chia}}, \bibinfo {author} {\bibfnamefont {H.}~\bibnamefont {Warner}},
  \bibinfo {author} {\bibfnamefont {B.}~\bibnamefont {Pingault}}, \bibinfo
  {author} {\bibfnamefont {B.}~\bibnamefont {Machielse}}, \bibinfo {author}
  {\bibfnamefont {S.}~\bibnamefont {Meesala}},\ and\ \bibinfo {author}
  {\bibfnamefont {M.}~\bibnamefont {Loncar}},\ }\href
  {https://arxiv.org/abs/2503.09946} {\bibinfo {title} {Observation of the
  acoustic {P}urcell effect with a color-center and a nanomechanical
  resonator}} (\bibinfo {year} {2025}),\ \Eprint
  {https://arxiv.org/abs/2503.09946} {arXiv:2503.09946 [quant-ph]} \BibitemShut
  {NoStop}%
\bibitem [{\citenamefont {Lee}\ \emph {et~al.}(2012)\citenamefont {Lee},
  \citenamefont {Singer},\ and\ \citenamefont {Thomas}}]{acoustic-meta}%
  \BibitemOpen
  \bibfield  {author} {\bibinfo {author} {\bibfnamefont {J.-H.}\ \bibnamefont
  {Lee}}, \bibinfo {author} {\bibfnamefont {J.~P.}\ \bibnamefont {Singer}},\
  and\ \bibinfo {author} {\bibfnamefont {E.~L.}\ \bibnamefont {Thomas}},\
  }\bibfield  {title} {\bibinfo {title} {Micro-/nanostructured mechanical
  metamaterials},\ }\href
  {https://doi.org/https://doi.org/10.1002/adma.201201644} {\bibfield
  {journal} {\bibinfo  {journal} {Advanced Materials}\ }\textbf {\bibinfo
  {volume} {24}},\ \bibinfo {pages} {4782} (\bibinfo {year}
  {2012})}\BibitemShut {NoStop}%
\bibitem [{\citenamefont {Dirac}(1930)}]{dirac}%
  \BibitemOpen
  \bibfield  {author} {\bibinfo {author} {\bibfnamefont {P.~A.~M.}\
  \bibnamefont {Dirac}},\ }\href@noop {} {\bibfield  {journal} {\bibinfo
  {journal} {Proc. Cambridge Phil. Soc.}\ }\textbf {\bibinfo {volume} {26}},\
  \bibinfo {pages} {376} (\bibinfo {year} {1930})}\BibitemShut {NoStop}%
\bibitem [{\citenamefont {Frenkel}(1934)}]{frenkel}%
  \BibitemOpen
  \bibfield  {author} {\bibinfo {author} {\bibfnamefont {J.}~\bibnamefont
  {Frenkel}},\ }\href@noop {} {\emph {\bibinfo {title} {Wave Mechanics}}}\
  (\bibinfo  {publisher} {Oxford University Press},\ \bibinfo {year} {1934})\
  p.\ \bibinfo {pages} {435}\BibitemShut {NoStop}%
\bibitem [{\citenamefont {Sundaresan}\ \emph {et~al.}(2015)\citenamefont
  {Sundaresan}, \citenamefont {Liu}, \citenamefont {Sadri}, \citenamefont
  {Sz\ifmmode~\mbox{\H{o}}\else \H{o}\fi{}cs}, \citenamefont {Underwood},
  \citenamefont {Malekakhlagh}, \citenamefont {T\"ureci},\ and\ \citenamefont
  {Houck}}]{mqrm}%
  \BibitemOpen
  \bibfield  {author} {\bibinfo {author} {\bibfnamefont {N.~M.}\ \bibnamefont
  {Sundaresan}}, \bibinfo {author} {\bibfnamefont {Y.}~\bibnamefont {Liu}},
  \bibinfo {author} {\bibfnamefont {D.}~\bibnamefont {Sadri}}, \bibinfo
  {author} {\bibfnamefont {L.~J.}\ \bibnamefont {Sz\ifmmode~\mbox{\H{o}}\else
  \H{o}\fi{}cs}}, \bibinfo {author} {\bibfnamefont {D.~L.}\ \bibnamefont
  {Underwood}}, \bibinfo {author} {\bibfnamefont {M.}~\bibnamefont
  {Malekakhlagh}}, \bibinfo {author} {\bibfnamefont {H.~E.}\ \bibnamefont
  {T\"ureci}},\ and\ \bibinfo {author} {\bibfnamefont {A.~A.}\ \bibnamefont
  {Houck}},\ }\bibfield  {title} {\bibinfo {title} {Beyond strong coupling in a
  multimode cavity},\ }\href {https://doi.org/10.1103/PhysRevX.5.021035}
  {\bibfield  {journal} {\bibinfo  {journal} {Phys. Rev. X}\ }\textbf {\bibinfo
  {volume} {5}},\ \bibinfo {pages} {021035} (\bibinfo {year}
  {2015})}\BibitemShut {NoStop}%
\bibitem [{\citenamefont {Gely}\ \emph {et~al.}(2017)\citenamefont {Gely},
  \citenamefont {Parra-Rodriguez}, \citenamefont {Bothner}, \citenamefont
  {Blanter}, \citenamefont {Bosman}, \citenamefont {Solano},\ and\
  \citenamefont {Steele}}]{rabi-model}%
  \BibitemOpen
  \bibfield  {author} {\bibinfo {author} {\bibfnamefont {M.~F.}\ \bibnamefont
  {Gely}}, \bibinfo {author} {\bibfnamefont {A.}~\bibnamefont
  {Parra-Rodriguez}}, \bibinfo {author} {\bibfnamefont {D.}~\bibnamefont
  {Bothner}}, \bibinfo {author} {\bibfnamefont {Y.~M.}\ \bibnamefont
  {Blanter}}, \bibinfo {author} {\bibfnamefont {S.~J.}\ \bibnamefont {Bosman}},
  \bibinfo {author} {\bibfnamefont {E.}~\bibnamefont {Solano}},\ and\ \bibinfo
  {author} {\bibfnamefont {G.~A.}\ \bibnamefont {Steele}},\ }\bibfield  {title}
  {\bibinfo {title} {Convergence of the multimode quantum {R}abi model of
  circuit quantum electrodynamics},\ }\href
  {https://doi.org/10.1103/PhysRevB.95.245115} {\bibfield  {journal} {\bibinfo
  {journal} {Phys. Rev. B}\ }\textbf {\bibinfo {volume} {95}},\ \bibinfo
  {pages} {245115} (\bibinfo {year} {2017})}\BibitemShut {NoStop}%
\bibitem [{\citenamefont {Clougherty}(2024)}]{dpc24}%
  \BibitemOpen
  \bibfield  {author} {\bibinfo {author} {\bibfnamefont {D.~P.}\ \bibnamefont
  {Clougherty}},\ }\bibfield  {title} {\bibinfo {title} {Variational approach
  to atom-membrane dynamics},\ }\href {https://doi.org/10.1063/5.0237141}
  {\bibfield  {journal} {\bibinfo  {journal} {APL Quantum}\ }\textbf {\bibinfo
  {volume} {1}},\ \bibinfo {pages} {046120} (\bibinfo {year}
  {2024})}\BibitemShut {NoStop}%
\bibitem [{\citenamefont {Clougherty}(2014)}]{dpc13}%
  \BibitemOpen
  \bibfield  {author} {\bibinfo {author} {\bibfnamefont {D.~P.}\ \bibnamefont
  {Clougherty}},\ }\bibfield  {title} {\bibinfo {title} {Quantum sticking of
  atoms on membranes},\ }\href {https://doi.org/10.1103/PhysRevB.90.245412}
  {\bibfield  {journal} {\bibinfo  {journal} {Phys. Rev. B}\ }\textbf {\bibinfo
  {volume} {90}},\ \bibinfo {pages} {245412} (\bibinfo {year}
  {2014})}\BibitemShut {NoStop}%
\bibitem [{\citenamefont {Sengupta}\ and\ \citenamefont
  {Clougherty}(2018)}]{sengupta}%
  \BibitemOpen
  \bibfield  {author} {\bibinfo {author} {\bibfnamefont {S.}~\bibnamefont
  {Sengupta}}\ and\ \bibinfo {author} {\bibfnamefont {D.~P.}\ \bibnamefont
  {Clougherty}},\ }\bibfield  {title} {\bibinfo {title} {Infrared problem in
  cold atom quantum physisorption on 2{D} materials},\ }\href@noop {}
  {\bibfield  {journal} {\bibinfo  {journal} {J. Phys.: Conf. Ser.}\ }\textbf
  {\bibinfo {volume} {1148}},\ \bibinfo {pages} {012007} (\bibinfo {year}
  {2018})}\BibitemShut {NoStop}%
\bibitem [{\citenamefont {Yu}\ \emph {et~al.}(1993)\citenamefont {Yu},
  \citenamefont {Doyle}, \citenamefont {Sandberg}, \citenamefont {Cesar},
  \citenamefont {Kleppner},\ and\ \citenamefont {Greytak}}]{yu93}%
  \BibitemOpen
  \bibfield  {author} {\bibinfo {author} {\bibfnamefont {I.~A.}\ \bibnamefont
  {Yu}}, \bibinfo {author} {\bibfnamefont {J.~M.}\ \bibnamefont {Doyle}},
  \bibinfo {author} {\bibfnamefont {J.~C.}\ \bibnamefont {Sandberg}}, \bibinfo
  {author} {\bibfnamefont {C.~L.}\ \bibnamefont {Cesar}}, \bibinfo {author}
  {\bibfnamefont {D.}~\bibnamefont {Kleppner}},\ and\ \bibinfo {author}
  {\bibfnamefont {T.~J.}\ \bibnamefont {Greytak}},\ }\bibfield  {title}
  {\bibinfo {title} {Evidence for universal quantum reflection of {Hydrogen}
  from liquid $^4${He}},\ }\href@noop {} {\bibfield  {journal} {\bibinfo
  {journal} {Phys. Rev. Lett.}\ }\textbf {\bibinfo {volume} {71}},\ \bibinfo
  {pages} {1589} (\bibinfo {year} {1993})}\BibitemShut {NoStop}%
\bibitem [{\citenamefont {Bunch}\ \emph {et~al.}(2008)\citenamefont {Bunch},
  \citenamefont {Verbridge}, \citenamefont {Alden}, \citenamefont {van~der
  Zande}, \citenamefont {Parpia}, \citenamefont {Craighead},\ and\
  \citenamefont {McEuen}}]{mceuen}%
  \BibitemOpen
  \bibfield  {author} {\bibinfo {author} {\bibfnamefont {J.~S.}\ \bibnamefont
  {Bunch}}, \bibinfo {author} {\bibfnamefont {S.~S.}\ \bibnamefont
  {Verbridge}}, \bibinfo {author} {\bibfnamefont {J.~S.}\ \bibnamefont
  {Alden}}, \bibinfo {author} {\bibfnamefont {A.~M.}\ \bibnamefont {van~der
  Zande}}, \bibinfo {author} {\bibfnamefont {J.~M.}\ \bibnamefont {Parpia}},
  \bibinfo {author} {\bibfnamefont {H.~G.}\ \bibnamefont {Craighead}},\ and\
  \bibinfo {author} {\bibfnamefont {P.~L.}\ \bibnamefont {McEuen}},\ }\bibfield
   {title} {\bibinfo {title} {Impermeable atomic membranes from graphene
  sheets},\ }\href {https://doi.org/10.1021/nl801457b} {\bibfield  {journal}
  {\bibinfo  {journal} {Nano Letters}\ }\textbf {\bibinfo {volume} {8}},\
  \bibinfo {pages} {2458} (\bibinfo {year} {2008})}\BibitemShut {NoStop}%
\bibitem [{\citenamefont {Clougherty}\ and\ \citenamefont
  {Kohn}(1992)}]{dpc92}%
  \BibitemOpen
  \bibfield  {author} {\bibinfo {author} {\bibfnamefont {D.~P.}\ \bibnamefont
  {Clougherty}}\ and\ \bibinfo {author} {\bibfnamefont {W.}~\bibnamefont
  {Kohn}},\ }\bibfield  {title} {\bibinfo {title} {Quantum theory of
  sticking},\ }\href@noop {} {\bibfield  {journal} {\bibinfo  {journal} {Phys.\
  Rev.\ B}\ }\textbf {\bibinfo {volume} {46}},\ \bibinfo {pages} {4921}
  (\bibinfo {year} {1992})}\BibitemShut {NoStop}%
\bibitem [{\citenamefont {Gradshteyn}\ and\ \citenamefont
  {Ryzhik}(1980)}]{g&r}%
  \BibitemOpen
  \bibfield  {author} {\bibinfo {author} {\bibfnamefont {I.~S.}\ \bibnamefont
  {Gradshteyn}}\ and\ \bibinfo {author} {\bibfnamefont {I.~M.}\ \bibnamefont
  {Ryzhik}},\ }\href@noop {} {\emph {\bibinfo {title} {Table of Integrals,
  Series, and Products}}}\ (\bibinfo  {publisher} {Academic Press},\ \bibinfo
  {year} {1980})\BibitemShut {NoStop}%
\bibitem [{\citenamefont {Koya}\ \emph {et~al.}(2021)\citenamefont {Koya},
  \citenamefont {Zhu}, \citenamefont {Ohannesian}, \citenamefont {Yanik},
  \citenamefont {Alabastri}, \citenamefont {Proietti~Zaccaria}, \citenamefont
  {Krahne}, \citenamefont {Shih},\ and\ \citenamefont {Garoli}}]{nanoporous}%
  \BibitemOpen
  \bibfield  {author} {\bibinfo {author} {\bibfnamefont {A.~N.}\ \bibnamefont
  {Koya}}, \bibinfo {author} {\bibfnamefont {X.}~\bibnamefont {Zhu}}, \bibinfo
  {author} {\bibfnamefont {N.}~\bibnamefont {Ohannesian}}, \bibinfo {author}
  {\bibfnamefont {A.~A.}\ \bibnamefont {Yanik}}, \bibinfo {author}
  {\bibfnamefont {A.}~\bibnamefont {Alabastri}}, \bibinfo {author}
  {\bibfnamefont {R.}~\bibnamefont {Proietti~Zaccaria}}, \bibinfo {author}
  {\bibfnamefont {R.}~\bibnamefont {Krahne}}, \bibinfo {author} {\bibfnamefont
  {W.-C.}\ \bibnamefont {Shih}},\ and\ \bibinfo {author} {\bibfnamefont
  {D.}~\bibnamefont {Garoli}},\ }\bibfield  {title} {\bibinfo {title}
  {Nanoporous metals: From plasmonic properties to applications in enhanced
  spectroscopy and photocatalysis},\ }\href
  {https://doi.org/10.1021/acsnano.0c10945} {\bibfield  {journal} {\bibinfo
  {journal} {ACS Nano}\ }\textbf {\bibinfo {volume} {15}},\ \bibinfo {pages}
  {6038} (\bibinfo {year} {2021})}\BibitemShut {NoStop}%
\bibitem [{\citenamefont {Wyatt}\ and\ \citenamefont {Klier}(2000)}]{puddles3}%
  \BibitemOpen
  \bibfield  {author} {\bibinfo {author} {\bibfnamefont {A.~F.~G.}\
  \bibnamefont {Wyatt}}\ and\ \bibinfo {author} {\bibfnamefont
  {J.}~\bibnamefont {Klier}},\ }\bibfield  {title} {\bibinfo {title} {Model for
  the extreme wetting hysteresis of liquid helium on cesium},\ }\href
  {https://doi.org/10.1103/PhysRevLett.85.2769} {\bibfield  {journal} {\bibinfo
   {journal} {Phys. Rev. Lett.}\ }\textbf {\bibinfo {volume} {85}},\ \bibinfo
  {pages} {2769} (\bibinfo {year} {2000})}\BibitemShut {NoStop}%
\bibitem [{\citenamefont {Ribeiro}\ \emph {et~al.}(2018)\citenamefont
  {Ribeiro}, \citenamefont {Martínez-Martínez}, \citenamefont {Du},
  \citenamefont {Campos-Gonzalez-Angulo},\ and\ \citenamefont
  {Yuen-Zhou}}]{polariton}%
  \BibitemOpen
  \bibfield  {author} {\bibinfo {author} {\bibfnamefont {R.~F.}\ \bibnamefont
  {Ribeiro}}, \bibinfo {author} {\bibfnamefont {L.~A.}\ \bibnamefont
  {Martínez-Martínez}}, \bibinfo {author} {\bibfnamefont {M.}~\bibnamefont
  {Du}}, \bibinfo {author} {\bibfnamefont {J.}~\bibnamefont
  {Campos-Gonzalez-Angulo}},\ and\ \bibinfo {author} {\bibfnamefont
  {J.}~\bibnamefont {Yuen-Zhou}},\ }\bibfield  {title} {\bibinfo {title}
  {Polariton chemistry: controlling molecular dynamics with optical cavities},\
  }\href {https://doi.org/10.1039/C8SC01043A} {\bibfield  {journal} {\bibinfo
  {journal} {Chem. Sci.}\ }\textbf {\bibinfo {volume} {9}},\ \bibinfo {pages}
  {6325} (\bibinfo {year} {2018})}\BibitemShut {NoStop}%
\bibitem [{\citenamefont {Surjadi}\ \emph {et~al.}(2019)\citenamefont
  {Surjadi}, \citenamefont {Gao}, \citenamefont {Du}, \citenamefont {Li},
  \citenamefont {Xiong}, \citenamefont {Fang},\ and\ \citenamefont
  {Lu}}]{mech-mat}%
  \BibitemOpen
  \bibfield  {author} {\bibinfo {author} {\bibfnamefont {J.~U.}\ \bibnamefont
  {Surjadi}}, \bibinfo {author} {\bibfnamefont {L.}~\bibnamefont {Gao}},
  \bibinfo {author} {\bibfnamefont {H.}~\bibnamefont {Du}}, \bibinfo {author}
  {\bibfnamefont {X.}~\bibnamefont {Li}}, \bibinfo {author} {\bibfnamefont
  {X.}~\bibnamefont {Xiong}}, \bibinfo {author} {\bibfnamefont {N.~X.}\
  \bibnamefont {Fang}},\ and\ \bibinfo {author} {\bibfnamefont
  {Y.}~\bibnamefont {Lu}},\ }\bibfield  {title} {\bibinfo {title} {Mechanical
  metamaterials and their engineering applications},\ }\href
  {https://doi.org/https://doi.org/10.1002/adem.201800864} {\bibfield
  {journal} {\bibinfo  {journal} {Advanced Engineering Materials}\ }\textbf
  {\bibinfo {volume} {21}},\ \bibinfo {pages} {1800864} (\bibinfo {year}
  {2019})}\BibitemShut {NoStop}%
\bibitem [{\citenamefont {Groth}\ \emph {et~al.}(2004)\citenamefont {Groth},
  \citenamefont {Kr{\"u}ger}, \citenamefont {Wildermuth}, \citenamefont
  {Folman}, \citenamefont {Fernholz}, \citenamefont {Mahalu}, \citenamefont
  {Bar-Joseph},\ and\ \citenamefont {Schmiedmayer}}]{chip04}%
  \BibitemOpen
  \bibfield  {author} {\bibinfo {author} {\bibfnamefont {S.}~\bibnamefont
  {Groth}}, \bibinfo {author} {\bibfnamefont {P.}~\bibnamefont {Kr{\"u}ger}},
  \bibinfo {author} {\bibfnamefont {S.}~\bibnamefont {Wildermuth}}, \bibinfo
  {author} {\bibfnamefont {R.}~\bibnamefont {Folman}}, \bibinfo {author}
  {\bibfnamefont {T.}~\bibnamefont {Fernholz}}, \bibinfo {author}
  {\bibfnamefont {D.}~\bibnamefont {Mahalu}}, \bibinfo {author} {\bibfnamefont
  {I.}~\bibnamefont {Bar-Joseph}},\ and\ \bibinfo {author} {\bibfnamefont
  {J.}~\bibnamefont {Schmiedmayer}},\ }\bibfield  {title} {\bibinfo {title}
  {Atom chips: Fabrication and thermal properties},\ }\href@noop {} {\bibfield
  {journal} {\bibinfo  {journal} {Appl.\ Phys.\ Lett.}\ }\textbf {\bibinfo
  {volume} {85}},\ \bibinfo {pages} {14} (\bibinfo {year} {2004})}\BibitemShut
  {NoStop}%
\bibitem [{\citenamefont {Sedlacek}\ \emph {et~al.}(2016)\citenamefont
  {Sedlacek}, \citenamefont {Kim}, \citenamefont {Rittenhouse}, \citenamefont
  {Weck}, \citenamefont {Sadeghpour},\ and\ \citenamefont
  {Shaffer}}]{sadeghpour}%
  \BibitemOpen
  \bibfield  {author} {\bibinfo {author} {\bibfnamefont {J.~A.}\ \bibnamefont
  {Sedlacek}}, \bibinfo {author} {\bibfnamefont {E.}~\bibnamefont {Kim}},
  \bibinfo {author} {\bibfnamefont {S.~T.}\ \bibnamefont {Rittenhouse}},
  \bibinfo {author} {\bibfnamefont {P.~F.}\ \bibnamefont {Weck}}, \bibinfo
  {author} {\bibfnamefont {H.~R.}\ \bibnamefont {Sadeghpour}},\ and\ \bibinfo
  {author} {\bibfnamefont {J.~P.}\ \bibnamefont {Shaffer}},\ }\bibfield
  {title} {\bibinfo {title} {Electric field cancellation on quartz by {R}b
  adsorbate-induced negative electron affinity},\ }\href
  {https://doi.org/10.1103/PhysRevLett.116.133201} {\bibfield  {journal}
  {\bibinfo  {journal} {Phys. Rev. Lett.}\ }\textbf {\bibinfo {volume} {116}},\
  \bibinfo {pages} {133201} (\bibinfo {year} {2016})}\BibitemShut {NoStop}%
\bibitem [{\citenamefont {Haroche}(2013)}]{haroche}%
  \BibitemOpen
  \bibfield  {author} {\bibinfo {author} {\bibfnamefont {S.}~\bibnamefont
  {Haroche}},\ }\bibfield  {title} {\bibinfo {title} {Nobel lecture:
  Controlling photons in a box and exploring the quantum to classical
  boundary},\ }\href {https://doi.org/10.1103/RevModPhys.85.1083} {\bibfield
  {journal} {\bibinfo  {journal} {Rev. Mod. Phys.}\ }\textbf {\bibinfo {volume}
  {85}},\ \bibinfo {pages} {1083} (\bibinfo {year} {2013})}\BibitemShut
  {NoStop}%
\end{thebibliography}%

\end{document}